\documentclass[reprint,aps,pra]{revtex4-1}

\usepackage{graphicx}
\usepackage{dcolumn}
\usepackage{bm}
\usepackage{amsmath}
\usepackage{amsthm}
\usepackage{amssymb}
\usepackage{float}
\usepackage{hyperref}
\usepackage{color}
\usepackage{epstopdf}
\usepackage{cleveref}
\usepackage[svgnames]{xcolor}
\hypersetup{hidelinks,colorlinks=true,allcolors=DarkBlue}

\begin{document}

\preprint{APS/123-QED}

\title{Gaussian mixture model for event recognition \\ in optical time-domain reflectometry based sensing systems}
\author{A.K. Fedorov$^{1,2}$}
\author{M.N. Anufriev$^{1,2}$}
\author{A.A. Zhirnov$^{1}$}
\author{K.V. Stepanov$^{1}$}
\author{E.T. Nesterov$^{1}$}
\author{D.E. Namiot$^{3}$}
\author{V.E. Karasik$^{1}$}
\author{A.B. Pnev$^{1}$}
\affiliation
{
	\mbox{$^{1}$Bauman Moscow State Technical University, 2nd Baumanskaya St. 5, Moscow 105005, Russia}
	\mbox{$^{2}$Theoretical Department, DEPHAN, Novaya St. 100, Skolkovo, Moscow 143025, Russia}
	\mbox{$^{3}$Lomonosov Moscow State University, Vorob'evy gory 1, Moscow 119992, Russia}
}

\date{\today}

\begin{abstract}
We propose a novel approach to the recognition of particular classes of non-conventional events in signals from phase-sensitive optical time-domain-reflectometry-based sensors. 
Our algorithmic solution has two main features: filtering aimed at the de-nosing of signals and a Gaussian mixture model to cluster them. 
We test the proposed algorithm using experimentally measured signals. 
The results show that two classes of events can be distinguished with the best-case recognition probability close to 0.9 at sufficient numbers of training samples. 
\end{abstract}
                              
\maketitle

The problem of the development of novel sensor techniques plays a crucial role in science and technology. 
One of the most important classes of sensing systems is distributed sensors, which are of great importance for the remote control of extended objects 
\cite{Bao,Feng,Fedorov,Kulchin,Taylor,Park,Choi,Gold}.
Phase-sensitive coherent optical time-domain reflectometry ($\Phi$--OTDR) is a basic technique that can provide sufficient sensitivity and resolution for these distributed sensing systems 
\cite{Taylor,Park,Choi}. 
Standard OTDR techniques use light sources with coherence lengths, which are shorter than pulse lengths. 
This can yield a sum of backscattered intensities from each scattering center, which allows one, {\it e.g.}, to control splices and breaks in fiber cables \cite{Gold}.
On the contrary, in  $\Phi$--OTDR-based sensing systems \cite{Juarez,Zhang,Rao,Lu,Qin,Martins}, 
the coherence length of lasers is longer than their pulse length. 
An event near the optical fiber generates an acoustic wave that affects the fiber by changing the phases of the backscattering centers. 
An analysis of such signals can reveal their impact on the sensor and monitor located near fiber objects~\cite{Juarez,Zhang,Rao,Lu,Qin,Martins}.
A key stage in implementing $\Phi$--OTDR-based sensors is the development of an algorithmic solution to reveal unusual vibrations in the background.

The problem of the recognition of non-conventional activity (a target event) consists of two closely related subproblems. 
The first is related to the de-noising procedure, which allows the detection of an event in the background with high probability. 
The second and much more important subproblem is the development of a classification methods aimed at clustering detected target events into predetermined classes.
Due to the complex structure of the signals in such sensors, this is challenging \cite{Qin2,Shi,Duan,Li,Fedorov2,Lyons,Sun,Wang,Wu}.

In addition to guaranteeing high accuracy of recognition, 
post-processing algorithms for $\Phi$--OTDR-based sensors should be able to operate rapidly without significant computational resources. 
In other cases, their application in real-time distributed fiber optic sensing systems is substantially limited. 
In vibration sensing systems based on $\Phi$--OTDR-based, there is at present no sufficiently fast and versatile algorithmic solution for the recognition of events.
A promising direction for the solution of this problem is using of a machine learning toolbox, in particular neural networks and pattern recognition techniques \cite{Lyons,Sun,Wu,Wang}. 
Recent results \cite{Sun,Wu,Wang} have shown that recognition algorithms based on machine learning yield a classification accuracy close to one with a reasonably small recognition time.

In this study, we employ techniques of machine learning for the recognition of target events in $\Phi$--OTDR-based distributed fiber-optic vibration sensor. 
The $\Phi$--OTDR based sensing setup was used to collect experimental data for this study (for details, see \cite{parameters,Pnev,Pnev2}). 
As a potential application of the system, we have in mind the problem of access control for protected regions, 
with the fiber being located along their perimeter. 
The system generates field distributions $I(l,\delta{t})$ corresponding to time interval $\delta{t}$.
These raw data are obtained from the setup in blocks with ${\sim}10^3$ reflectograms, 
which are the result of measurements in one second. 
The input for post-processing is the block $I_t(l)$ containing ${\sim}10^3$ reflectograms.

The proposed detection procedure commences with the de-noising of measured signals $I_t(l)$ by the median filter in the time domain of the dimension three.
The procedure is intended to reduce the self-noises of the system. 
The structure of the self-noise resembles that of random noise, for the removal of which median filtering is an effective technique \cite{Huang}.

In the next stage, we use another type of filtering. 
This is based on the $3$--$700$ Hz band-pass filter to remove the DC component of the noise as well as noise at high frequencies. 
The boundaries for the band-pass filter are adaptive, {\it i.e.}, they have been optimized based on an analysis of the experimental data. 
Thus, we obtain the de-noised signal $X(l,t)$.

Following filtering, for every point-space step $l_j$, we calculate the standard deviation with a window of  $\sim20$--$50$ ms. 
The parameters of the time window were optimized according to the set of experimental data. 
However, the set of data with  $\sigma(l_j)$ for every $j$ was very large. 
Hence, we used a thinning procedure, where we split vector  $\sigma(l_j)$ on a region of 100 counts, calculate the maximum value for these regions, 
and form a new vector $\sigma^{t}(l_j)$  consisting of these maximum values.

Then, we can detect events based on whether threshold $X_{\rm c}$ is exceeded. 
This value is calculated adaptively with respect to the concrete set of experimental data.
In fact, following the employment of the procedure described above, the detection of potential target events reduces to a search for local centers of mass in the space--time field distribution $X(l,t)$,
which exceeds the critical value. 
For our specific set of experimental data, the critical value could be chosen as $10$--$100$ of the standard deviation of reflectograms in the last 15 minutes. 
For our case, this critical value was $X_{\rm c}\simeq10^6$. 
It was calculated on the basis of an analysis of the instantaneous values of the signals, 
which guaranteed events, and that of the mean values of low noise points (only technical self-noises). 
That is, signals that did not contain non-conventional events and had a minimal effect on seismic noise had mean values of $\bar{X}\simeq10^3$ and standard deviations of $\sigma_X\simeq10^3$, 
and those signifying high noise points (seismic noises on top of self-noises) had mean values of  $\bar{X}_{\rm noise}\simeq10^5$ and $\sigma_{\rm X\, noise}\simeq10^4$. 
This yields efficient detection. 

For an event, we allocate a region of the signal containing it and other events in its vicinity in the time domain.
We are interested in the case where we have to separate two specific classes of events: single-target passage, and digging near the cable. 

On the one hand, it is important to note that our detection procedure allows us to efficiently locate events. 
On the other, important properties that can serve as benchmarks for recognition are lost during this procedure. 
That is why we separate the stages, and why input data for classification is once again constituted by raw measured reflectograms $I_t(l)$.

The first step is the formation of a feature space for the machine learning tool aimed at classification of detected events. 
For this purpose, we use cepstral coefficients \cite{Huang2}, which are used in machine learning toolboxes aimed on recognition of complex acoustic signals \cite{Shaughnessy,GMM}. 
Here, we briefly describe their calculation for signal $I_t(l)$ of discrete form in time domain $I_l[t]$, where $0\le{t}<N$.
It is important to note that for the formation of the feature space, we used a raw space--time field distribution.

We apply to the signal $I_l[t]$ the Fourier transform 
\begin{equation}\label{Fourier}
	Y_l[k]=\sum\nolimits_{t=0}^{N-1}{I_l[t]\exp\left[-{2\pi{i}}kt/N\right]}, \quad 0\le{k}<N.
\end{equation}
Using the window function, we employ the following sequence of filters 
\begin{equation}\label{window}
	\mathcal{H}_m(k)=
	\left\{
	\begin{array}{ll}
	0, \qquad\qquad\qquad k<f\left[m-1\right],
	\\
	\frac{k-f\left[m-1\right]}{f\left[m\right]-f\left[m-1\right]}, \quad f\left[m-1\right]\leq{k}<f\left[m\right],
	\\
	\frac{f\left[m+1\right]-k}{f\left[m+1\right]-f\left[m\right]}, \quad f\left[m\right]\leq{k}\leq{f\left[m+1\right]},
	\\
	0, \qquad\qquad\qquad k>f\left[m-1\right],
    \end{array}\right.
\end{equation}
where the frequency $f\left[m\right]$ can be obtained from the following relation
\begin{equation}
	f\left[m\right]=\frac{N}{F_S}\left[B^{-1}\left(B(f_1)+\frac{B(f_n)-B(f_1)}{M+1}\right)\right],
\end{equation}
Here, $M$ is the number of Mel coefficients, $F_s{=}f_d/2{=}1$ kHz is the Nyquist sampling rate, $f_d$ is the sampling frequency, and $B(x)$ is the transformation to the Mel scale:
$B^{-1}(x){=}700(\exp\left[{x}/{1125}\right]-1)$,
where the numerical coefficient corresponds to the most common formula for transformation to the Mel scale \cite{Shaughnessy}.

We then calculate the energy for every window (\ref{window})
\begin{equation}\label{energy}
	S\left[m\right]=\ln\left(\sum\nolimits_{k=0}^{N-1}\left|Y_l[k]\right|^2\mathcal{H}_m(k)\right).
\end{equation}
By applying the discrete cosine transformation to (\ref{energy}),
\begin{equation}
	c\left[n\right]=\sum\nolimits_{m=0}^{M-1}S\left[m\right]\cos\left(\frac{\pi{n}(m+1/2)}{M}\right),  
\end{equation}
with $0\leq{n}<M$, 
we obtain the set of the cepstral coefficients in the Mel scale.  

The second step of the algorithm is the assignment of an event to classes. 
We use the Gaussian mixture model (GMM) for this. 
The GMM is a probabilistic model. 
The main assumption underlying the GMM is that all data points are generated from a mixture of a finite number of Gaussian distributions with unknown parameters. 
In our case, there are two set of data points with a Gaussian distribution. 
The feature space for the GMM is formed by the cepstral coefficients. We found that the optimal value of the cepstral coefficients was $M=10$.

The workflow of the GMM is as follows: 
The input data for classification is a vector $\mathbf{K}$ of cepstral coefficients of the signal. 
In the GMM framework \cite{GMM}, 
$\mathbf{K}$ is interpreted as representing independent observations from a mixture of two multivariate normal distributions which, in fact, form the feature space for classification. 
Two normal distributions are then generated, 
where $\mathcal{N}(\mu_1,\Sigma_1)$  corresponds to the input vector belonging to the first class and $\mathcal{N}(\mu_2,\Sigma_2)$ corresponds to that belonging to the second class. 
Furthermore, the goal is to estimate unknown parameters representing the ``mixing'' value between these normal distributions.
After this, the GMM works on the basis of the expectation--maximization algorithm \cite{GMM2}.
Thus, the algorithm computes for each point a probability of being generated by each component of the model.
Consequently, one can tweak the parameters to maximize the likelihood of the data given those assignments. 
Repeating this process guarantees convergence to a local optimum. In terms of implementation, we used the GMM library described in the details in Ref. \cite{GMM3}.

\begin{table}\label{tab:results}
\begin{tabular}{||c|c|c|c|c||}
\hline
Tr. samples & Test. samples & Correct & Correct, \% \\
\hline 
$35 $&$141$&$103$&$65,96$\\
$44$&$132$&$97$&$73,48$\\
$52$&$124$&$103$&$83,06$\\
$61$&$115$&$93$&$80,87$\\
$70$&$106$&$91$&$85,85$\\
$79$&$97$&$77$&$79,38$\\
$88$&$88$&$82$&$93,18$\\
$96$&$80$&$66$&$82,5$\\
$105$&$71$&$63$&$88,73$\\
$114$&$62$&$56$&$90,32$\\
$123$&$53$&$50$&$94,34$\\
$132$&$44$&$38$&$86,36$\\
$140$&$36$&$32$&$88,89$\\
\hline
\end{tabular}
\caption
{
Parameters of the test for the experimental verification of the proposed algorithm: 
number of training samples (tr.~samples) for the GMM, number of test samples (test.~samples), number of correctly clustered events and  percentage. 
}
\label{table:learning}
\end{table}

The collection of experimental data was organized as follows: 
We are interested in two types of events: single-target passage, and digging near the cable. 
During the experiment, we obtained 3.93 GB of raw data (2.51 GB corresponded to digging samples, and 1.42 GB to passage samples). 
We then collected 58 single-target passages at a distance of $0$--$5$ m from the cable and 118 digging events at $0$--$10$ m from it under similar conditions.
The suggested algorithmic solution was executed on a standard PC with an Intel Core i5 M 460 processor and 8 GB of RAM. 
The average time taken by the classification stage of the algorithm was approximately $1$ ms, which is reasonable in light of recently reported results. 
Therefore, the proposed algorithm is sufficiently fast to implement in real-time distributed fiber optic vibration sensing systems to control protected areas.

The estimated signal-to-noise ratio (SNR) in the collected experimental data was approximately  $8.5$ dB \cite{comment4}.
The effect of changes in SNR in the region on recognition was sufficiently small in this case. 
The probability of recognition, which is defined as the ratio of successful recognition of events to all input events, with a varying number of training samples 
({\it i.e.}, samples used for the formation of clusters) and testing samples is listed in Table I. 
Our results showed that the two classes of events can be detected and distinguished with the best-case probability close to 0.9 using the proposed method.
However, as it clearly seen from the table the probability strongly depends on number and properties of testing samples. 
These issues are interesting to be studied in future. 

The authors thank the anonymous referees for their useful comments. 
This work was partially supported by the Ministry for Education and Science of the Russian Federation within the Federal Program under Contract 14.579.21.0104 (A.K.F. and M.N.A.).


\begin{thebibliography}{9}
\bibitem{parameters}
We use a semiconductor laser with an erbium-doped fiber amplifier with power limited to $1$~W.
The wavelength of the probe signal was $1550$ nm, 
duration of the probe pulse was $100$--$500$ ns, 
coherence length of the laser is 30 km.
and the signal was launched into a standard single-mode telecommunication optical fiber SMF28 of length approximately $50$ km.
The fiber is probed on ranges of a few km.
The use of multimode fibers is not needed for our purposes because it affects spatial resolution.
\bibitem{comment5}
Since the recognition algorithm assigns to the class input events only to existed classes, 
it is important to exclude the effect of different types, 
which is beyond the classes considered here. 
For this, one needs to define through a preliminary test a probability threshold $p_c$. 
If the algorithm predicts that an input event belongs to the first class with probability $p_1$, 
this event belongs to the second type with probability $p_2=1-p_1$. 
The workflow of the algorithm stops in cases where one of these probabilities is less than the threshold.
\end{thebibliography}
\end{document}